\documentclass[prl,aps,twocolumn,floats,showpacs,epsf]{revtex4}
\usepackage{amssymb}

\usepackage{epsfig}

\newcommand{\be}{\begin{equation}}
\newcommand{\ee}{\end{equation}}
\newcommand{\bea}{\begin{eqnarray}}
\newcommand{\eea}{\end{eqnarray}}

\newcommand{\p}{\partial}

\newcommand{\la}{\langle}
\newcommand{\ra}{\rangle}
\newcommand{\rd}{\mbox{d}}
\newcommand{\ri}{\mbox{i}}
\newcommand{\re}{\mbox{e}}

\begin{document}
\title{Phenomenological theory of the underdoped phase of a
 high-T$_c$  superconductor.}

\author{A.  M. Tsvelik$^{1}$  and A. V. Chubukov$^{2}$}
\affiliation{$^{1}$ 
 Department of  Condensed Matter Physics and Materials Science, Brookhaven National Laboratory, Upton, NY 11973-5000, USA}
\affiliation{$^{2}$Department of University of Wisconsin, Madison, WI 53706, USA}
\date{\today}

\begin{abstract}
We model the Fermi surface of the cuprates by
 one-dimensional  nested parts near $(0,\pi)$ and $(\pi,0)$
 and unnested parts near the 
zone diagonals. Fermions in the nested regions  form 1D spin liquids, and
 develop spectral gaps below some $\sim T^*$, but superconducting order is 
 prevented by 1D phase fluctuations. 
 We show that the  Josephson  coupling between
  order parameters at  $(0,\pi)$ and $(\pi,0)$ 
 locks their relative phase at a crossover scale $T^{**}< T^*$. Below $T^{**}$, the  
system response becomes two-dimensional, and
 the system  displays Nernst effect. The remaining 
total phase gets locked at $T_c < T^{**}$, at which 
 the   system develops a (quasi-) long-range superconducting order.
\end{abstract}

\maketitle

 The copper oxide materials in their underdoped phase combine features typical for systems of different dimensionality. The existence of 
stripes~\cite{stripes,Kivelson}, the absence of quasiparticle peaks 
 above $T_c$,  the  presence of the spectral gap near  near $(\pi,0)$ and $(0,\pi)$ at $T > T_c$, and the filling of the gap by thermal fluctuations 
(instead of closing of the gap)~\cite{timusk,norman}, are hallmarks of 1D strongly correlated systems. 
On the other hand, the $d$-wave form of the pairing gap and  
existence  of curved portions of the Fermi surface (FS) near zone diagonals  
~\cite{ARPES} are  essentially 
two-dimensional features 
responsible for the physics of superconducting cuprates at the lowest energies 
(see \cite{chubukov}).  An intriguing aspect of the cuprate physics is the existence of the
 region, above $T_c$, where the system displays a strong diamagnetic response~\cite{ong}.  This region extends to some $T^{**}$ which is 
 smaller than the pseudogap temperature $T^*$, and has a different doping dependence. 

 In this letter we  propose a semi-phenomenological model which 
proposes an interpretation of the origin of the difference between $T_c$, $T^{**}$ and $T^*$. 
  We assume that fermions in antinodal regions near $(0,\pi)$ and $(\pi,0)$ interact 
attractively and form quasi-one-dimensional spin liquids (SL) 
with a large spin gap. These electrons then
 generate low energy collective modes whose dispersion is highly anisotropic in space. This assumption is consistent with  with ARPES experiments~\cite{ARPES} showing incoherent spectral functions  near the antinodes.  The assumption of strong anisotropy of collective modes is also
  consistent with the `checkerboard''-like density 
modulation seen in the STM experiments \cite{davis}, and with the recently observed softening of the optical phonons at wave-vectors  $(4k_F,0)$
 and  $(0,4k_F)$ (Ref. \cite{reznik}).

 The effective interaction in spin liquids is attractive and scales to strong coupling. At the other hand,
 it is reasonable to suggest that the coupling at the unnested parts and between the unnested and nested parts remains  weak and repulsive. The increase in doping leads to a decrease in the anisotropy which  explains opposite trends in doping dependence of $T^{*}$ and $T^{**}$. Without discussing any  microscopic details  of the 1D SL,  we just exploit 
 the known fact that   
interactions there open up  quasiparticle gaps
 (pseudogaps)    below some
 crossover temperature $T^*$, which is the largest energy scale in  our consideration. In 3D, the gap opening would imply spin or charge ordering, in 1D 
the ordering is prevented by strong  quantum phase fluctuations.

We consider the  Josephson-type coupling between the two
 fluctuating superconducting 
 order parameters formed near  $(0,\pi)$ and $(\pi,0)$ and demonstrate that 
below a crossover temperature $T^{**} < T^*$, 
 fluctuations of the relative phase of the 
two  order parameters at $(0,\pi)$ and $(\pi,0)$
acquire a finite mass $m \sim T^{**}$, i.e., the 
  relative phase gets locked. This implies that at $T^{**}$ 
the system crosses over into a 2D fluctuational regime with enhanced diamagnetic response, and begins displaying Nernst effect.  We further show that
 due to topological excitations (vortices) in the 2D regime, the (quasi)-long-range superconducting order develops only at a smaller $T_c < T^{**}$, where the system undergoes XY-model type phase transition. 
Below $T_c$ the $d$-wave-like spectral gap opens on the  entire FS  due to the proximity effect. 

\begin{figure}
\begin{center}
\epsfxsize=0.25\textwidth
\epsfbox{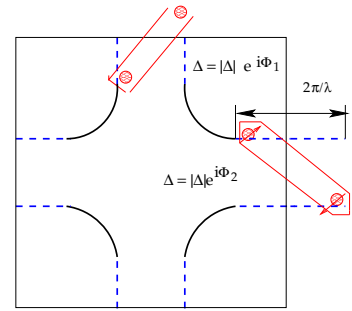}
\end{center}
\caption{A sketch of the Fermi surface with pairing. Spectral gaps open on the nested (blue) regions.}
\label{net}
\end{figure}
Our approach bears some similarities  to earlier theories, 
 but there are important differences. 
The ``stripe'' scenario~\cite{Kivelson} also departs from 1D physics, but it assumes the existence of two different 1D subsystems, separated in  real space,
one with a spin gap, and one without. In this situation, the the dominant interaction is within each 
subsystem. In our scenario, the 1D SLs near $(0,\pi)$ and $(\pi,0)$ are identical, both having a spin gap, and the dominant interaction is between these two SLs. 
Besides, in stripe scenario, the nodal
 quasiparticles are produced by the interaction between the stripes, while we obtain nodal quasiparticles due to the interaction between 1D and curved parts of the FS.
  In the  RVB scenario\cite{RVB},\cite{rice}  SL is supposed to be two-dimensional, and strong phase fluctuations at $T=0$ emerge from the fact that the (Hubbard) interaction is assumed to be larger than the fermionic bandwidth unlike in our model where strong phase fluctuations are due to quasi-1D behavior.  In  ``quantum-critical'' scenarios~\cite{adv,grilli},
 the strongest interaction, responsible for $T^*$, is also assumed to be within the antinodal region, and the $d-$wave 
 superconducting gap emerges due to the residual interaction between the antinodal regions and the rest of the FS.
 However, in these scenarios, the FS is assumed to be almost circular, and the quasi-1D behavior is absent.
%

In our model, the  FS, shown in Fig. 1, has nested regions, and   
 the system can  be separated into two quasi-1D systems represented by the
 electronic states close to the antinodal regions, and a 2D
 system represented by the states close to the unnested part of the Fermi surface.
 The low energy physics is dominated by   
 two gapless collective modes originating from  critical fluctuations of the 
charge in 1D regions, and by nodal quasiparticles. 
Each critical mode is very anisotropic dispersing either in the $x$ or $y$ direction. As is always the case in 1D, such modes 
  describe simultaneously fluctuations of  the charge density wave (CDW) 
 $\Delta_{CDW}$ and the superconducting  (SC) 
 $\Delta_{SC} = \Delta$ order parameter . 
The corresponding phases $\Theta_x$ and $X$ are  
 related to each other by duality $
\p_{\mu}X = \epsilon_{\mu\nu}\p_{\nu}\Theta_x, ~~ (\mu,\nu = v\tau, x)
$, and the 1D effective action can 
be equally well expressed in terms of $X$ or $\Theta_x$. Below we use 
$\Delta_{x,y}$ as the basic variables (subindex refers to  fluctuations near $(0,\pi)$ and $(\pi,0)$, respectively), and also switch from $\Delta_{x,y}$ 
to the phases $X,Y$ using $
\Delta_x = |\Delta|\re^{\ri X}, ~~ \Delta_y = |\Delta|\re^{\ri Y}$. 
At energies  $E << |\Delta| \sim T^*$, 
 the order parameter amplitudes are frozen, but the phases fluctuate.  
We assume that the most relevant interaction between fermions 
 near $(0, \pi)$ and $(\pi,0)$ is the Josephson coupling between the phases of the two SC order parameters $\Delta_x$ and $\Delta_y$ 
 (note that the CDW phases $\Theta_{x,y}$ cannot be coupled due to the momentum conservation).

The dynamics of the order parameters $X$ and $Y$ 
 is described by   the effective action in the form of coupled 1D XY models:
 $S = \int^{1/T} d \tau \int\frac{dx dy}{\lambda} 
{\cal L}_1 $, where $ {\cal L}_1 $ is 
\bea
&& \frac{\rho}{2} [(\dot X)^2 + (\dot Y)^2] - \frac{t}{{\lambda}^{2}}\cos[X(x,y) - X(x+\lambda,y)]\label{1D}\\
&& -\frac{t}{{\lambda}^{2}}\cos[Y(x,y) - Y(x,y +\lambda)] + J\cos[X(x,y) - 
Y(x,y)] \nonumber
\eea
Here $2\pi/\lambda$ is the size of the nested area of the Fermi surface along a particular  direction (experimentally $\lambda \approx 4a$, where $a$ is the lattice constant).  For later convenience we discretized the gradient term in the action replacing it by the cosine. 
The Coulomb interaction can be absorbed into the 
model parameters as it is screened by the ungapped  nodal 
quasiparticles.   
The last term in (\ref{1D}) 
describes the Josephson coupling between the 
two SC order parameters.  
We assume that this coupling is repulsive, i.e., 
$J > 0$, in which case it favors  $X-Y \approx \pi$.  

At $J = 0$ the action (\ref{1D}) describes critical and purely 1D fluctuations of the phases $X,Y$. Parameters of the model can be expressed in terms of 
collective mode velocity $v$ and the scaling dimension $d$ of the SC order parameter 
\be
\rho = (4\pi dv)^{-1}, ~~ t = v/4\pi d
\ee
which can be extracted from the order parameter correlation 
function, e.g.,  $\chi_{xx}  = \la\la\re^{\ri X(\tau,x,y)}\re^{-\ri X(0,0,y)}\ra\ra$.
\bea
\chi_{xx}  = \left[\frac{(\pi T)^2}{\sin^2(\pi T\tau) + \sinh^2(\pi Tx/v)}\right]^{d}\label{corrchi}
\eea
($\chi_{yy}$ is obtained by interchanging of $x$ and $y$).

Deviations from one-dimensionality near $(0,\pi)$ or $(\pi,0)$ are taken into account by the  terms of the form
\begin{eqnarray}
&&{\cal L}_{2} \sim
b_{\Phi} (\cos[X(x,y) - X(x,y + \lambda)] + \nonumber \\
&& \cos[Y(x,y) - Y(x + \lambda,y)]) + (X,Y \rightarrow 4\pi\Theta_{x,y} d)
\end{eqnarray}
These  extra terms by itself 
lead to a finite temperature phase transition to a superconducting  or a CDW state (depending on what type of coupling is more relevant), even at 
$J =0$~\cite{Kivelson}. We assume, however, 
 that the corresponding transition temperature is the smallest energy scale in the problem, and neglect it, focusing  instead on the Josephson interaction.
 
The coupling of the SC order parameters to the unnested part of the Fermi surface is given by 
\bea
[\gamma(k_x)\Delta_x(q) + \gamma(k_y)\Delta_y(q)]\psi^+_{-k + q/2}\psi^+_{k + q/2} + H.c. \label{int}
\eea
where $\gamma(...)$ are phenomenological parameters.
 This interaction opens the  SC gap on the curved parts of FS through the proximity effect when the phases of $\Delta_x$ and $\Delta_y$ lock. At $k_x = k_y$, the  couplings $\gamma$ are 
equal, and for $X - Y = \pi$ (i.e.,  $\Delta_x = - \Delta_y$)
  the interaction vanishes.  
Setting $\gamma(k_x) = \gamma\cos k_x$, we then 
reproduce the $d$-wave order parameter along the whole FS.
When the phases $X,Y$ are 
not locked, the interaction (\ref{int})  is unable to generate a gap at the unnested part of FS, but it can destroy the quasiparticle coherence along the whole Fermi surface. Indeed,  evaluating the self energy of these  electrons from (\ref{int},\ref{corrchi}) in the leading order in $\gamma$ we obtained that it is local and behaves as
\bea
Im \Sigma \sim \gamma^2\Delta^2\int \rd\tau \frac{\sin(\omega\tau)}{\tau} |\tau|^{-2d} \sim \omega^{2d}
\eea
It can be shown
 that the higher order terms are less singular in $\omega$. If $d < 1$ this self-energy exceeds $\omega^2$ coming from other,  non-singular interactions. 
 Experimentally, $\Sigma (\omega)$ scales nearly linearly with $\omega$ above $T_c$. 
In our phenomenological theory, this 
implies that $d \approx 1/2$. 

We now return to  effective action 
(\ref{1D}). Assuming that $X,Y (x,y)$  are slowly varying functions of $x$ and $y$ respectively and replacing $t/{\lambda}^2$ terms in (\ref{1D}) by quadratic functions, we obtain the action  quadratic in $\Phi^{(+)} = (X + Y)/\sqrt 2$. After integrating out $\Phi^{(+)}$ we obtain the action for 
$\Phi^{-} = (X - Y)/\sqrt 2$ in the form
$S_{eff} =  \int \frac{d \omega d k_x dk_y}{(2\pi)^3} {\cal L}_{eff} (k, \omega)$
where 
\bea
&& {\cal L}_{eff} (k, \omega) =
\frac{1}{2} 
\Phi^{(-)}[G^{-1}_0(\omega,k_x) + G^{-1}_0(\omega,k_y)]\Phi^{(-)} + \nonumber\\
&& 
J (2\pi)^3 \delta ({\bf k}) \delta (\omega) 
\left[\cos(\sqrt 2\Phi^{(-)})\right]_{k, \omega}, \label{Model}
\eea
where  $G_0(\omega,k) = 2 \lambda d[(\omega^2/v + vk^2)]^{-1} $ and $[..]_{k,\omega}$ means Fourier-transform.

 Models of this kind  have been discussed  before in the context of sliding Luttinger liquid phases \cite{lub}.  It was  assumed there 
that the theory is renormalizable. By inspecting the  perturbation theory series in $g$ we have  found 
 that this is not the case. In this situation,  
arguments based on a supposed  relevance of various operators, used
 in \cite{lub} are questionable. In our analysis,
 we restrict ourselves to the summation of the most diverging 
 RPA
diagrams. 
The RPA series for the pairing susceptibility 
 are given by
\bea
\chi_{xx}(\omega;{\bf k}) = \{[\chi_0(\omega,k_x)]^{-1} - J^2\chi_0(\omega,k_y)\}^{-1}, \label{RPA1}
\eea
where $\chi_0(\omega,k)$ is the Fourier transform of (\ref{corrchi}).
Evaluating $\chi_0 (k, \omega)$ (see Ref. \cite{efetov}) 
and substituting into (\ref{RPA1}) we find that
 $\chi_{xx}(0,{\bf 0})$ diverges at the temperature 
\be
T^{**} = \frac{\Lambda}{2\pi}\left(\frac{J\lambda}{\Lambda}\sin\pi d \frac{\Gamma^2(d/2)\Gamma^2(1 -d)}{\Gamma^2(1 -d/2)}\right)^{\frac{1}{2 -2d}}\label{RPA}
\ee
where
 $\Lambda = 2\pi v/\lambda$. The divergence implies, that within RPA, 
$T^{**}$ is  the true superconducting ordering temperature~\cite{comm}.  Below $T^{**}$,
 phase fluctuations acquire a finite mass $m(T)$. The spectrum  can be obtained by extending RPA analysis to the ordered state, i.e., by replacing  cosine Josephson 
potential by a quadratic function of ${\tilde \Phi}^{-} = \sqrt 2 \Phi^{(-)} - \pi$ as $ J\cos[\sqrt 2 \Phi^{(-)}] \approx -\frac{m^2}{2} \left({\tilde \Phi}^{-}\right)^2$
\bea
  m^2 = J|\la\cos[\sqrt 2 \Phi^{(-)}]\ra| =  J\exp\left[-\frac{1}{2}\la \left({\tilde \Phi}^{-}\right)^2\ra\right].
\label{ac2}
\eea
Plugging this back into 
(\ref{Model}),
 we get 
\bea
&& \la\la{\tilde \Phi}_i(-\omega, -k) {\tilde \Phi}_j(\omega, k)\ra\ra = \label{ac1}\\
&& \frac{1}{D(\omega,{\bf k})}\left(
\begin{array}{cc}
\rho\omega^2 + tk_y^2 + m^2 & m^2\\
m^2 & \rho\omega^2 + tk_x^2 + m^2
\end{array}
\right)
\nonumber
\eea
where 
$i,j = x,y$,
 $D = (\rho\omega^2 + tk_y^2)(\rho\omega^2 + tk_x^2) + m^2(2\rho\omega^2 + t{\bf k}^2)$ 
Its poles at $\ri\omega = E_{\pm}$ yield the  excitation spectrum:
\be
E^2_{\pm} = \frac{k^2}{2} + (m/2)^2 \pm \sqrt{(m/2)^4 + \frac{1}{4}k^4\cos^2(2\theta)}
\ee
where $k_x = k\cos\theta, k_y = k\sin\theta$. 
The mode $E_-$ is gapless, as it should be in the ordered state.
A substitution of (\ref{ac1}) into (\ref{ac2})  yields the 
self-consistent  equation for $m$ solving which  we reproduce (\ref{RPA}). 

As we shall see,the vortices  transform $T^{**}$ into
 a crossover scale at which  fluctuations of $\Phi^{-}$ develop a gap, and the system 
 response becomes 2D.  The static diamagnetic susceptibility, however,  
 still remains finite, i.e., the system does not become a superconductor. 
To demonstrate this, we use the Villain approximation replacing all cosine potentials in the effective action (\ref{1D}) by {\it periodic} quadratic functions,
 e.g.,  
\bea
&& -\cos[X(x,y) - X(x +\lambda,y)] \rightarrow \\
&& \frac{1}{2}[X(x,y) - X(x +\lambda,y) - 2\pi N_1(x,x+\lambda;y)]^2\nonumber
\eea
We further represent $
N_1(x,x+\lambda;y) = \tilde N_1(x,y) - \tilde N_1(x+\lambda,y) + N_1(x,x+\lambda;y)$.
The integers $N_{1,2}$ live on links of the lattice, while 
$\tilde N_{1,2}$ live on its sites.  One can easily verify that the 
actions for $\tilde N_{1,2}$ and $N_{1,2}$ decouple from each other. 
 Because of the space limitations, we derive  only the action for 
 $\tilde N_{1,2}$. These fields  enter $S_{eff}$ 
in combination  $\tilde N_1 - \tilde N_2 \equiv n$. The Lagrangian density  is 
\bea
&& {\cal L} = \rho[(\dot X)^2 + (\dot Y)^2] + t(\p_x X)^2 + t(\p_y Y)^2 + \nonumber\\
&& m^2[X(x,y) - Y(x,y) - 2\pi n(x,y)]^2
\label{ac_n}
\eea
Integrating over $X,Y$ fields we obtain the effective action for $n$:
\bea
S[n] = 2\pi^2 m^2 T\sum_{\omega, {\bf k}}n(-\omega, -{\bf k})K(\omega, {\bf k})n(\omega, {\bf k})
\eea
\bea
K(\omega, {\bf k}) =\frac{(\rho\omega^2 + tk_x^2)(\rho\omega^2 + tk_y^2)}{(\rho\omega^2 + tk_x^2)(\rho\omega^2 + tk_y^2) + m^2(2\rho\omega^2 + t{\bf k}^2)}\nonumber
\eea
At finite $T$ the most important contribution comes from zero frequency where the kernel is 
\bea
K(0, {\bf k}) = tk_x^2k_y^2/(tk_x^2k_y^2 + m^2{\bf k}^2) \label{kernel}
\eea
This form of the kernel  suggests that the relevant integer variable is $
q(x,y) = \p_x\p_y n(x,y)$ 
so that $
n(x,y) = \sum_j e_j\theta(x - x_j)\sum_k e'_k\theta(y - y_k)$, where 
$e_j,e'_k$ are integers. 
Thus from S[n] we arrive to  the 
classical action for the Coulomb gas of  ``charges'' $q(j,k) = e_je'_k$:
\bea
S[q] = \frac{2\pi t}{T}\sum_{{\bf r},{\bf r}'}q_{\bf r}\ln\left[|{\bf r} -{\bf r}'|/b\right] q_{{\bf r}'}, ~~ b \sim m^{-1} \label{Coul}
\eea
This form implies that the actual ordering transition 
belongs to the universality class of XY model (2D one for a single layer or anisotropic 3D one for a system of weakly coupled layers). 
The transition temperature then can be estimated as
\be
T_{c} = \frac{\pi}{2} \frac{\pi t}{\lambda} = \frac{\pi v}{8 d \lambda} = \frac{\Lambda}{16 d} 
\label{ac_n1}
\ee
 Below $T_c$ the vortices are bound into pairs, both the relative and total phase of the two order parameters are locked, 
and the system displays  Meissner effect. The coupling to the unnested part of the FS then gives rise to the emergence of the $d-$wave order parameter all along the FS. 
 By construction, the phases X, Y have to be locked first, hence $T_c$ must be smaller than $T^**$. Parametrically, this requires $J$ to be quite large:
 $J \lambda > \Lambda = 2\pi v/\lambda \sim (\pi/2) v/a$.
 However, numberwise, for $d=1/2$, we obtain from 
(\ref{RPA}) and (\ref{ac_n1}) $T^**/T_c \approx 70 (J \lambda/\Lambda)$, hence 
 $T^** > T_c$ even if $J \lambda \leq \Lambda$.
%

The analysis of the effective action for integer fields $N_{1,2}$ 
proceeds in a similar fashion.
After the integration over $X,Y$ we obtain
\bea
 S = 2\pi^2 t \sum (N_1, N_2) {\hat K} \left(\begin{array}{c}
N_1\\
N_2
\end{array}
\right)\nonumber
\eea
The kernel matrix ${\hat K} $ is such that at zero frequency we have the same Coulomb gas action (\ref{Coul}) (with the same $T_c$!) for the charges  $
q = \p_yN_1 - \p_x N_2$.  In the presence of a magnetic field 
the link variables are modified to
 $N_1 \rightarrow 2e \lambda A_x/c + N_1$ and $N_2 \rightarrow 2e \lambda A_y/c + N_2$, and the charges acquire an extra piece $q \rightarrow q + \frac{2e {\lambda}^2B}{c}$,
where $B$ is the magnetic field. This implies that these charges are directly coupled to a magnetic field, i.e.,  between $T^{**}$ and $T_{c}$ 
 the system develops a strong diamagnetic response 
and  displays Nernst effect. 

 The existence of the diamagnetic response in a finite range above 
$T_c$ is  consistent with the measurements of the Nernst effect in the cuprates~\cite{ong}. In agreement with our theory the temperature up to which Nernst effect has been observed is definitely much smaller than $T^*$. 
 Experimentally, the onset temperature for the Nernst effect follows the same trend as $T_c$ and goes down 
in strongly underdoped cuprates implying that  the ratio $J\lambda/\Lambda$ is independent on doping. Our theory also predicts that at $T > T_c$,
the interaction between nested and curved parts of the FS 
 destroys fermionic coherence all along the Fermi surface. 

 To summarize, in this paper we proposed a phenomenological model 
for the low energy physics of underdoped cuprates.  
We assumed that at some large energy scale  
$T^*$, fermions within  the nested regions (see Fig.) 
 form 1D spin liquids.  
 Strong quantum 
1D  fluctuations prevent the superconductivity to develop at $T^*$; these fluctuations are supressed only 
at much lower temperature $T^{**}$
where the relative phase of the two order parameters  made of  electronic states near  $(0,\pi)$ and $(\pi,0)$ 
gets locked, 
 and the system  crosses over into a 2D fluctuational regime 
with enhanced diamagnetic response.   Because of the topological vortex 
fluctuations 
 the total phase of the two order parameters
 becomes locked  only at even smaller  $T_{c} < T^{**}$, below which the system develops a 
(quasi-) long-range superconducting order.

AVC acknowledges the support from Theory Institute for Strongly Correlated and Complex Systems at BNL and NSF DMR 0240238.  AMT  is grateful to Abdus Salam ICTP for hospitality and acknowledges the support from
 US DOE under contract number DE-AC02 -98 CH 10886. We are grateful to A. A. Nersesyan for valuable ideas,
and to  A. G. Abanov, D. Basov, G. Blumberg, E. Fradkin, A. Parmekanti, T. M. Rice,  O. Starykh and J. Tranquada
 for discussions and interest to the work.

\end{document}